\documentclass[conference]{IEEEtran}
\IEEEoverridecommandlockouts

\usepackage{cite}
\usepackage{amsmath,amssymb,amsfonts}
\usepackage{graphicx}
\usepackage{textcomp,makecell}
\usepackage[table]{xcolor}
\usepackage{booktabs,svg,pgfplots,tikz,tikzscale,subcaption,adjustbox}
\usetikzlibrary{calc,positioning,patterns}
\usepgfplotslibrary{external} 
\pgfplotsset{compat=newest}
\newcounter{colend}
\newcounter{maxcol}
\newcounter{mincol}

\newcounter{facnt}
\newcounter{hacnt}
\newcounter{dotcnt}

\newcommand*\drawdotcol[3][0]{
  \setcounter{colend}{#3}
  \ifnum \value{colend} > 0 {   
    \addtocounter{colend}{-1}%
    \addtocounter{colend}{#1}%
    \foreach \t in {#1,...,\value{colend}}{
      \draw[fill=black] (\value{maxcol}-#2,\t) circle (0.2);
      \addtocounter{dotcnt}{1}
    }
  }
  \else {
    \draw[opacity=0,fill=black] (\value{maxcol}-#2,0) circle (0.2);
  }
  \fi
}

\newcommand*\drawtrunccol[3][0]{ 
  \setcounter{colend}{#3}
  \addtocounter{colend}{-1}%
  \addtocounter{colend}{#1}%

  \ifnum #3 > 0   
    \foreach \t in {#1,...,\value{colend}}{
      \draw[black!50, fill=white] (\value{maxcol}-#2,\t) circle (0.2);
  	  \addtocounter{dotcnt}{1}
    }
  \fi
}

\usepackage{rotating}
\usepackage[hidelinks]{hyperref}

\usepackage{balance}

\usepackage{multirow}
\usepackage{array}
\newcolumntype{C}[1]{>{\centering\arraybackslash}p{#1}} 
\newcolumntype{R}[1]{>{\raggedleft\arraybackslash}p{#1}} 
\newcolumntype{L}[1]{>{\raggedright\arraybackslash}p{#1}} 

\usepackage{bm}

\begin{document}
\title{Small Logic-based Multipliers with\\Incomplete Sub-Multipliers for FPGAs}

\author{\IEEEauthorblockN{Andreas Böttcher, Martin Kumm}
\IEEEauthorblockA{\textit{Fulda, University of Applied Sciences, Faculty of Applied Computer Science}, Fulda, Germany \\
andreas.boettcher@cs.hs-fulda.de, martin.kumm@cs.hs-fulda.de}
}

\maketitle

\begin{abstract}
There is a recent trend in artificial intelligence (AI) inference towards lower precision data formats down to 8 bits and less.
As multiplication is the most complex operation in typical inference tasks, there is a large demand for efficient small multipliers.
The large DSP blocks have limitations implementing many small multipliers efficiently.
Hence, this work proposes a solution for better logic-based multipliers that is especially beneficial for small multipliers.
Our work is based on the multiplier tiling method in which a multiplier is designed out of several sub-multiplier tiles.
The key observation we made is that these sub-multipliers do not necessarily have to perform a complete (rectangular) $\bm N$\null$\times$\null$\bm K$ multiplication and more efficient sub-multipliers are possible that are incomplete (non-rectangular).
This proposal first seeks to identify efficient incomplete irregular sub-multipliers and then demonstrates improvements over state-of-the-art designs. 
It is shown that optimal solutions can be found using integer linear programming (ILP), which are evaluated in FPGA synthesis experiments. 
\end{abstract}


\section{Introduction}
With multiplication being one of the most pervasive operations in computer arithmetic, its efficient implementation is of ongoing interest. When targeting FPGAs the design methods have to consider the embedded DSP blocks and look-up-tables (LUTs) as the primary resources of modern devices. DSP blocks are usually the most convenient to use, as they allow the implementation of a multiplication by themselves (for example up to one 27$\times$27 or two 18$\times$18 signed or unsigned on Intel Stratix10 or 25$\times$18 signed or 24$\times$17 unsigned on AMD 7 Series) but are a relatively limited resource. Multiplier design methods have to accommodate for that to avoid the under-utilization or premature depletion of DSP resources when generating complex circuits \cite{Dinechin09,Banescu10,Kumm15,Kumm17}. A multitude of methods have been proposed to remedy the underlying design challenge \cite{Dinechin09,Banescu10,ParandehAfshar11,Brunie13,Walters14,Kumm15,Walters16,fwsakw17,Kumm17,Langhammer19,lb18,Langhammer19DSP,Langhammer20DSP,Boettcher20}.

In \cite{fwsakw17,Langhammer19DSP,Langhammer20DSP,Xilinx2020} the extraction of smaller multipliers from one DSP by additional LUT-based circuitry was proposed. Conversely, when larger multipliers than the size of a single DSP are needed, Karatsuba's algorithm can be used to reduce the number of DSPs \cite{Dinechin09, Boettcher20}. With Slice LUTs being one of the most prolific resources on FPGAs, several concepts that efficiently use LUT resources in conjunction with the fast-carry-chain like implementations of vendor-specific Booth-Arrays \cite{Walters14, Kumm15, Walters16}, Baugh-Wooley-Multipliers  \cite{ParandehAfshar11} and several optimizations with the fractal-synthesis approach \cite{Langhammer19,lb18} were devised.     

Another design process is multiplier tiling \cite{Dinechin09} which seeks to achieve the efficient use of DSPs by combining them with supplementary LUT-based sub-multipliers. But as aside from heuristics \cite{Banescu10,Brunie13,Boettcher20} also optimal \cite{Kumm17,Boettcher21} methods for solving the underlying tiling problem are known, in the sense that for a given set of sub-multipliers (tiles) the combination with the least realization cost in terms of logic resources is found while constraining the use of DSPs to a maximal number. Hence, to generate smaller designs as in the present proposal, where DSPs would be considerably underutilized, by constraining the number of DSPs to zero, multiplier tiling can be used to generate logic-only solutions which constitute the optimal use of LUT resources. Multipliers generated with tiling usually consist of the sub-multipliers (tiles) as partial product generators, which are subsequently compressed on a compressor tree to form a single output vector. The compressor tree design for FPGAs has to consider certain peculiarities in contrast to ASICs, as it has to be implemented within the confines of the considered target architecture. As full- or half-adders under-utilize the FPGA slice hardware, more efficient structures like generalized parallel counters (GPCs)  \cite{Meo75,ParandehAfshar08,ParandehAfshar08b,ParandehAfshar09,ParandehAfshar11c,Kumm2014EfficientHS,Kumm14,Preusser17,Yuan19}, 4:2 row compressors \cite{Kumm2014EfficientHS,Hormigo09,Kamp09} and the ternary adder \cite{sp06,blsy09} were devised, which utilize the fast-carry-chain of modern FPGAs. The introduction of the additional compressors significantly extended the design space to be explored to find efficient solutions, which necessitates advanced concepts for compressor tree design beyond Dadda \cite{Dadda65}. 
Several heuristic \cite{ParandehAfshar08,ParandehAfshar09,ParandehAfshar11c} and optimal \cite{ParandehAfshar08b,Matsunaga11,Matsunaga13,Kumm14,kk18,Boettcher23} methods have been presented for this task. Since the selection of the partial product generators (tiles), compressor tree and final adder design are interdependent problems, as a different choice in one of the steps can result in an alternative combination becoming the the optimal solution in another, a recent combined approach \cite{Boettcher23} that optimizes both the tiling and the compressor tree simultaneously, is used in the present proposal.

Except for the Karatsuba algorithm with rectangular DSPs \cite{Kumm18} and DSP super tiles \cite{Dinechin09} the previous mapping schemes usually aimed on creating rectangular (sub-)multipliers.

In AI inference tasks, it is commonly accepted that 8 bit integers are sufficient to reach floating point accuracy with a proper scaling and quantization-aware training in many applications.
Recent transformer networks are typically quantized to even 4 bits. This motivates this work to focus on small multipliers.

The present proposal %
\begin{itemize}
\item presents a novel class of incomplete logic-based (sub-) multipliers that map efficiently to most modern FPGAs,
\item demonstrates the integration and use in conjunction with the multiplier tiling methodology to design complete multipliers, 
\item demonstrates that the proposed designs yield a reduction in utilized resources compared to other state-of-the-art logic-based multiplier design methods while maintaining similar properties in terms of critical path delay and latency.
\end{itemize}

\section{Multiplier Tiling}
Consider a large $2W_X\times 2W_Y$ sized multiplier. The $X \times Y$ multiplication operation can be decomposed into the weighted sum of four $W_X\times W_Y$ sub-multiplications $M_{1,..,4}$ according to 
\begin{align}
\label{eq:schoolbook_tiling}
X \times Y&=\left(x_h 2^{W_X}+x_l\right)\left(y_h 2^{W_Y}+y_l\right) \notag \\
&=\underbrace{x_h y_h}_{M_4}2^{W_X+W_Y}+\underbrace{x_h y_l}_{M_3}2^{W_X}+\underbrace{x_l y_h}_{M_2}2^{W_Y}+\underbrace{x_l y_l}_{M_1} \ . \
\end{align}
Thereby the input vectors $X$ and $Y$ are split into a most and least significant sub-word $x_{h},x_{l}$ and $y_{h},y_{l}$ with a respective size of $W_X$ and $W_Y$.
Accordingly by splitting the input vectors $X$ and $Y$ into the individual bits, this process can be repeated down to the respective partial product bits, as they would be generated by the AND-gates of a Bough-Wooley multiplier. However, in practical tilings usually larger multipliers are used which map more efficiently to the hardware of the particular FPGA.
This process is considered graphically with the multiplier tiling methodology, such that the multiplier to be realized is represented as a rectangular $2W_X\times 2W_Y$ sized board which is subdivided into a grid according to the bits of the input vectors $X,Y$, where each position corresponds to a partial product bit of an AND-array-multiplier.

\begin{figure}
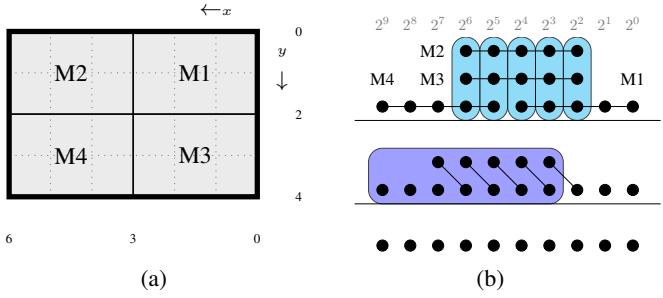

\centering
\begin{subfigure}[b]{0.23\textwidth}
    \centering
    \includegraphics[width=1\textwidth]{./figs/6x4tiling_example.tikz}
    \caption{}
    \label{fig:6x6tiling}
\end{subfigure}
\begin{subfigure}[b]{0.25\textwidth}
    \centering
    \includegraphics[width=1\textwidth]{./figs/6x4compression_example.tikz}
    \caption{}
    \label{fig:6x6compression}
\end{subfigure}
\caption{Tiling (a) and compressor tree (b) of a $6\times 4$-multiplier, composed of four $3\times 2$-tiles.}
\label{fig:multiplier_architecture}
\end{figure}

\figurename~\ref{fig:multiplier_architecture} shows an example of a $6\times 4$ multiplier, assembled from four $3\times 2$ sub-multiplier tiles  $M_{1,..,4}$. Each of the  $3\times 2$-tiles generates a 5-bit output vector which subsequently has to by processed on the compressor tree to calculate the output vector of the multiplier as shown in \figurename~\ref{fig:6x6compression}.

\subsection{Conventional Rectangular Multiplier Tiles}

\begin{table}[t]
\caption{Properties of LUT- and DSP-based multipliers \cite{Boettcher20} 
targeting AMD FPGAs, where $\text{cost}^\text{mult}_e$ and $\text{cost}^\text{tile}_t$ are the number of LUT6 for the single multiplier and multiplier+compression, respectively
}
\centering
\setlength\tabcolsep{3pt}
\begin{tabular}{ c c c c c c l c}
\toprule
Type & Shape & $A_t$ & $\text{cost}^\text{mult}_t$ & $\text{cost}^\text{tile}_t$ & $W_\text{out}$ & $E_t$ & $D^t$ \\
\midrule
d    & 1$\times$1                   &   1   & 1     & 1.65  & 1             & 0.625 & 0\\
c  & 1$\times$2 / 2$\times$1      &   2   & 1     & 2.3   & 2             & 0.87  & 0\\
b  & 2$\times$3 / 3$\times$2      &   6   & 3     & 6.25  & 5             & 0.96  & 0\\
a    & 3$\times$3                   &   9   & 5     & 8.9   & 6             & 1.011 & 0\\
e  & 2$\times$\null$k$ / $k$\null$\times$2      &  2$k$ & $k$+1 & 1.65$k$ + 2.3 & $k$+2 & $\frac{2k}{1.65k + 2.3}$    & 0\\
\cmidrule(rl){1-8}
    -& 24$\times$17 / 24$\times$17  &  408  & 0     & 26.65          & 41& 15.30 &1\\
\bottomrule
\end{tabular}
\label{table:tiles_properties}
\end{table}

\def \tilescale{1}
\def \tilescaleouter{0.22}
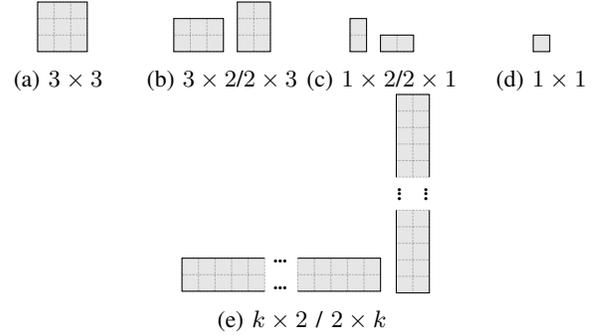
\begin{figure}[!t]
	\centering
	\subcaptionbox{$3\times 3$ \label{fig:logictile1}}[2.0cm]{
	\scalebox{\tilescaleouter}{
	\begin{tikzpicture}[black,scale=\tilescale]
      \draw[line width=1pt,fill=black!10] (0,0) -- ++ (3, 0) -- ++ (0,3) -- ++ (-3,0) -- cycle;
      \draw[thin,dotted] grid (3,3);
	\end{tikzpicture}
	}
}
\subcaptionbox{$3\times 2$/$2\times 3$\label{fig:logictile2}}[2.0cm]{
  \scalebox{\tilescaleouter}{
  \begin{tikzpicture}[black,scale=\tilescale]
      \draw[line width=1pt,fill=black!10] (0,0) -- ++ (3, 0) -- ++ (0,2) -- ++ (-3,0) -- cycle;
      \draw[thin,dotted] grid (3,2);
	\end{tikzpicture}	
  }
  \scalebox{\tilescaleouter}{
	\begin{tikzpicture}[black,scale=\tilescale]
      \draw[line width=1pt,fill=black!10] (0,0) -- ++ (2, 0) -- ++ (0,3) -- ++ (-2,0) -- cycle;
      \draw[thin,dotted] grid (2,3);
  \end{tikzpicture}  
  }
}
\subcaptionbox{$1\times 2$/$2\times 1$\label{fig:logictile3}}[2.0cm]{
  \scalebox{\tilescaleouter}{
	\begin{tikzpicture}[black,scale=\tilescale]
      \draw[line width=1pt,fill=black!10] (0,0) -- ++ (1, 0) -- ++ (0,2) -- ++ (-1,0) -- cycle;
      \draw[thin,dotted] grid (1,2);
  \end{tikzpicture}  
  }
  \scalebox{\tilescaleouter}{
  \begin{tikzpicture}[black,scale=\tilescale]
      \draw[line width=1pt,fill=black!10] (0,0) -- ++ (2, 0) -- ++ (0,1) -- ++ (-2,0) -- cycle;
      \draw[thin,dotted] grid (2,1);
	\end{tikzpicture}	
  }
}
\subcaptionbox{$1\times 1$\label{fig:logictile4}}[2.0cm]{
  \scalebox{\tilescaleouter}{
	\begin{tikzpicture}[black,scale=\tilescale]
      \draw[line width=1pt,fill=black!10] (0,0) -- (1, 0) -- (1,1) -- (0,1) -- cycle;
      \draw[thin,dotted] grid (1,1);
  \end{tikzpicture}  
  }
}
\subcaptionbox{$k\times 2$ / $2\times k$\label{fig:logictile5}}{
  \scalebox{\tilescaleouter}{
  \begin{tikzpicture}[black,scale=\tilescale]
      \draw[line width=1pt,fill=black!10] (5,2) -- (0,2) -- (0,0) -- (5, 0);
      \draw (6.3,1.6)  node {\scalebox{2}{\Huge \begin{rotate}{90}{\vdots}\end{rotate}}};
      \draw (6.3,0)  node {\scalebox{2}{\Huge \begin{rotate}{90}{\vdots}\end{rotate}}};
      \draw[line width=1pt,fill=black!10] (7,2) -- (12,2) -- (12,0) -- (7, 0);
      \draw[thin,dotted] grid (5,2);
      \draw[thin,dotted] (7,0) grid (12,2);
  \end{tikzpicture}  
  }
  \scalebox{\tilescaleouter}{
  \begin{tikzpicture}[black,scale=\tilescale]
      \draw[line width=1pt,fill=black!10] (2,5) -- (2,0) -- (0,0) -- (0,5);
      \draw (1.8,6.2)  node {\scalebox{2}{\Huge \vdots}};
      \draw (0.2,6.2)  node {\scalebox{2}{\Huge \vdots}};
      \draw[line width=1pt,fill=black!10] (2,7) -- (2,12) -- (0,12) -- (0,7);
      \draw[thin,dotted] grid (2,5);
      \draw[thin,dotted] (0,7) grid (2,12);
  \end{tikzpicture}  
  }
}
	\caption{Geometric shapes of rectangular LUT-based tiles\cite{Boettcher20}}
	\label{fig:logictiles}
\end{figure}

The previous multiplier tiling based design methods \cite{Kumm17,Boettcher20,Boettcher21,Boettcher23} used, aside from DSP tiles, a set of logic-based sub-multiplier tiles listed in Table~\ref{table:tiles_properties} and illustrated in \figurename~\ref{fig:logictiles}. Those tiles possess several properties relevant for the tiling algorithm. First of all, the tiles have a particular shape defined by the input vectors $X$ and $Y$.
The previously used LUT-based multipliers usually had a rectangular shape. Associated with the shape, the tiles cover a particular number of positions on the board which defines their geometric area $A_t$. Furthermore, the implementation of the sub-multiplier tiles has certain realization costs in terms of 6-input look-up-tables defined by $\text{cost}^\text{mult}_t$. Each tile also generates a $W_\text{out}$ wide output vector of partial product bits calculated by the individual LUTs. The partial product bits also entail additional costs for the calculation of the final result on the compressor tree, which had been shown in \cite{Kumm17} experimentally to be about $0.65$ LUTs per bit. So the cost of utilizing a particular tile $\text{cost}^\text{tile}_t$ can be described as,
\begin{equation}
    \text{cost}^\text{tile}_t=\text{cost}^\text{mult}_t+\widetilde{\text{cost}}^\text{comp}_t \ 
\end{equation}
where $\widetilde{\text{cost}}^\text{comp}_t$ is the compression cost of approximately $0.65\frac{\text{LUT}}{\text{Bit}}\cdot W_\text{out}^t$. This can further be used to derive a measure of efficiency $E_t$ for a particular tile as the quotient of the geometric area covered by the tile divided by its realization cost (Area per LUT)
\begin{equation}
    E_t=\frac{A_t}{\text{cost}^\text{tile}_t} .
\end{equation}
The efficiency of the tiles is provided for reference purposes to benchmark the properties of the tile, whereas the utilized combined optimization of the tiling and compression uses the exact number of LUTs required for compression.

Tiles like the embedded DSPs as shown in the bottom row of Table~\ref{table:tiles_properties} can also possess DSP costs denoted by $D^t$ according to the number of utilized DSPs. Note that the DSP block do not have any LUT cost $\text{cost}^\text{mult}_t$ on their own but cause compression costs, so they are listed with $\text{cost}^\text{tile}_t=26.65$LUT assuming complete utilization. When underutilized the cost has to be calculated dynamically.

\subsection{Motivational Example}

\begin{figure}[!t]
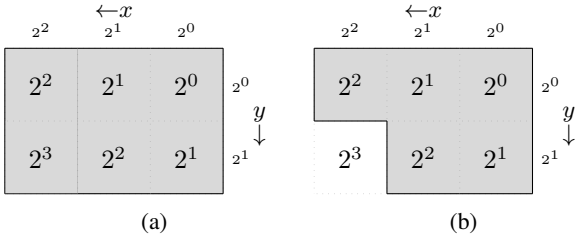

\centering
\begin{subfigure}[b]{0.22\textwidth}
    \centering
    \includegraphics[width=1\textwidth]{./figs/3x2tile.tikz}
    \caption{}
    \label{fig:3x2tile}
\end{subfigure}
\begin{subfigure}[b]{0.22\textwidth}
    \centering
    \includegraphics[width=1\textwidth]{./figs/pp_omittance.tikz}
    \caption{}
    \label{fig:3x2tilie_missing}
\end{subfigure}
    \caption{Motivational example}
	\label{fig:mot_example}
\end{figure}

Consider the rectangular $3\times 2$ tile as illustrated in \figurename~\ref{fig:mot_example}a. 
The maximal output value in the unsigned case is $(2^3-1) \times (2^2-1) = 21$, so the required word size is $W_\text{out}^t=5$ bit.
Together with the $2+3=5$ input bits, the tile is realized using five 5-input LUTs (in the following denoted as 5-LUTs).
The 6-LUTs of recent FPGAs can be subdivided into two 5-LUTs, so, three 6-LUTs are necessary to implement this tile. 
Note that one of the 6-LUTs remains underutilized as it is only occupied by one 5-LUT. 
The other 5-LUT is typically wasted as all its inputs are already fixed.
The overall cost of the tile is
\begin{equation}
    \text{cost}^\text{tile}_t=\text{cost}^\text{mult}_t+0.65\cdot W_\text{out}^t=3+3.25=6.25 \ 
\end{equation}
and the efficiency with a total geometric area of $A_t=6$ is 
\begin{equation}
    E_t=\frac{6}{6.25}=0.96 . 
\end{equation}

The maximal output value can be alternatively computed by summing the weights shown in \figurename~\ref{fig:logictiles}a, i.e., $2^0+2^1+2^1+2^2+2^2+2^3=21$. 
Now consider how the properties of the tile would change if the most significant partial product bit with weight $2^3$ would be omitted (\figurename~\ref{fig:mot_example}b). 
As a consequence, the maximum output value is reduced to $21-8=13$, thus, requiring only $W_\text{out}^t=4$ output bits.
This also leads to a better utilizing of $14$ out of $16$ possible values (88\%) instead of the 22 out of 32 values (69\%) of the $3\times 2$ tile.

Accordingly, only four 5-LUTs are needed to realize the tile, which can efficiently be mapped to two 6-LUTs. This corresponds to total realization costs of 
\begin{equation}
    \text{cost}^\text{tile}_t=2+2.6=4.6 \ 
\end{equation}
and an efficiency with the area reduced by one to $A_t=5$ is 
\begin{equation}
    E_t=\frac{5}{4.6}=1.087. 
\end{equation}
It can be observed that the omission of one the partial product bits leads to a new tile variant, which is more efficient then the previously known LUT tiles, and hence allows for tilings which require less LUTs per area.
Note that in terms of efficiency the same could be achieved by removing the least significant partial product bit with weight $2^0$ at the top right corner, which also necessitates the inclusion of an additional bit for a single position, that can only be mapped as an underutilized LUT. In this case the omitted LSB of the tile is effectively always zero.
Without context those structures could appear useless as they do not compute a complete multiplication, but as a tile it can be used to form larger complete multipliers. This observations raise the question which other efficient structures are possible.

\section{Searching efficient structures}
\input{figs/4x4search_results_over1}
This section covers the realization of a complete search for efficient structures.
As elaborated previously, multiplier tiles can realize random partial products within a larger multiplier, which can be represented by positions on the board. This defines the applicable design space to be explored. Application specific multipliers can be required in random sizes, but the search space has to be constructed to manageable dimensions. The search space has to incorporate all viable combinations that could be mapped within the 6-input LUTs present on most modern FPGAs. The largest rectangular multiplier without the use of the fast-carry-chain is the $3\times3$, which is constrained by the six available inputs of the LUTs on which the equation of a particular partial product bit may depend on. Hence for example, a $4\times2$ or $5\times1$ multiplier would also be possible but is less efficient in terms of covered area per LUT. Also geometrically stretched out versions of the multipliers with gaps between the utilized input variables should be possible within the search space, as they can be realized with otherwise the same properties. This defines the evaluated search space, as combinations beyond this would constitute redundant, even more stretched out versions of other tiles. A geometric bound of $4\times4$ was used here, which results in $2^{16}$ possible combinations to be evaluated.

\subsection{Implementation of the Search}
To enumerate the evaluated $4\times 4$ search space, as each partial product on the board can either be present or absent, the area was abstracted as a $16$-bit binary number. As this includes a considerable number of redundant patterns, each combination not geometrically aligned with active partial products on the top and right of the board are discarded. The evaluation of the patterns is well parallelizable, as it only depends on the respective pattern. So multiple value ranges can be investigated in concurrent tasks. Based on the partial product pattern and the signedness of the input vectors the output bits of the tile can be tabulated as a function of all possible binary combinations of the inputs. From the truth table a set of equations for the tile outputs can be created. The equations were subsequently simplified with the Quine-McClusky method to remove redundant input-dependencies. 

The necessity for the simplification is best explained by using an example. Take the truth table of the $3\times3$ tile in \figurename~\ref{fig:logictiles}a: it depends on six input bits and produces $3+3=6$ output bits. Hence, a naive implementation would required six 6-LUTs. However, the simplified equations for the two output LSBs results in
\begin{align}
r_0 &=x_0y_0\\
r_1 &=x_1y_0 + x_0y_1 \ .
\end{align}
Hence, they only depend on four variables that can be mapped into a single 6-LUT (realizing two 5-LUTs).

Within the resulting equations the number of utilized inputs can be counted to determine the optimal LUT-mapping by evaluating the possible combinations to subdivide a $6$-LUT into two $5$-LUTs, or placing an equation with 7 input dependencies into two 6-LUTs, etc. So the efficiency of a particular tile in terms of the utilized LUTs in relation the the covered area can be determined to insert the tile to a linked list to hold the results according to its efficiency level.

\def \tilescale{1}
\def \tilescaleouter{0.4}

\begin{figure}[!t]
    \centering
    \begin{subfigure}[b]{0.23\textwidth}
        \centering
        \scalebox{\tilescaleouter}{
        \begin{tikzpicture}[black,scale=\tilescale]
        	\draw[step=1.0,dotted,thin, black] (0,0) grid (6,6);
         \fill[fill=orange,fill opacity=0.3] (0, 0) rectangle (2, 2);
         \fill[fill=orange,fill opacity=0.3] (4, 4) rectangle (6, 6);
         \fill[fill=orange,fill opacity=0.3] (4, 0) rectangle (6, 2);
         \fill[fill=orange,fill opacity=0.3] (0, 4) rectangle (2, 6);
          \fill[fill=green,fill opacity=0.3] (2, 4) rectangle (4, 6);
          \fill[fill=green,fill opacity=0.3] (0, 2) rectangle (2, 4);
          \fill[fill=green,fill opacity=0.3] (3, 3) rectangle (6, 4);
          \fill[fill=green,fill opacity=0.3] (4, 2) rectangle (6, 3);
          \fill[fill=green,fill opacity=0.3] (2, 0) rectangle (3, 2);
          \fill[fill=green,fill opacity=0.3] (3, 0) rectangle (4, 1);
        	\fill[fill=magenta,fill opacity=0.5] (3, 4) rectangle (2, 3);
        	\draw (3, 4) -- (3 ,3);
        	\draw (3, 4) -- (2, 4);
        	\draw (2, 4) -- (2, 3);
        	\fill[fill=magenta,fill opacity=0.5] (3, 3) rectangle (2, 2);
        	\draw (2, 3) -- (2, 2);
        	\draw (3, 2) -- (2, 2);
        	\fill[fill=magenta,fill opacity=0.5] (4, 3) rectangle (3, 2);
        	\draw (4, 3) -- (4 ,2);
        	\draw (4, 3) -- (3, 3);
        	\fill[fill=magenta,fill opacity=0.5] (4, 2) rectangle (3, 1);
        	\draw (4, 2) -- (4 ,1);
        	\draw (3, 2) -- (3, 1);
        	\draw (4, 1) -- (3, 1);
        \fill[pattern=north west lines, pattern color=green] (3, 0) rectangle (4, 1);
        \fill[pattern=north west lines, pattern color=green] (0, 4) rectangle (2, 3);
        \fill[pattern=north west lines, pattern color=green] (1, 3) rectangle (2, 2);
        \foreach \x in {0,...,5}{
		      \node[gray, above] at (6-0.5-\x,6) {\Large $x_{\x}$};
            \node[gray, right] at (6.5, 6-\x,) {\Large $y_{\x}$};
	    }
        \draw (7,0.8)  node {$\downarrow$};
        \draw (7,1.2)  node {$y$};
        \draw (1.0,6.8)  node {$\leftarrow$ $x$};
        \end{tikzpicture}
        }
        \caption{}
        \label{fig:s_shaped_vert}
    \end{subfigure}
    \begin{subfigure}[b]{0.25\textwidth}
        \centering
        \scalebox{\tilescaleouter}{
        \begin{tikzpicture}[black,scale=\tilescale]
        	\draw[step=1.0,dotted,thin, black] (0,0) grid (6,6);
         \fill[fill=orange,fill opacity=0.3] (0, 0) rectangle (2, 2);
         \fill[fill=orange,fill opacity=0.3] (4, 4) rectangle (6, 6);
         \fill[fill=orange,fill opacity=0.3] (4, 0) rectangle (6, 2);
         \fill[fill=orange,fill opacity=0.3] (0, 4) rectangle (2, 6);
          \fill[fill=green,fill opacity=0.3] (2, 4) rectangle (4, 6);
          \fill[fill=green,fill opacity=0.3] (0, 2) rectangle (2, 3);
          \fill[fill=green,fill opacity=0.3] (0, 3) rectangle (1, 4);
          \fill[fill=green,fill opacity=0.3] (3, 3) rectangle (6, 4);
          \fill[fill=green,fill opacity=0.3] (4, 2) rectangle (6, 3);
          \fill[fill=green,fill opacity=0.3] (2, 0) rectangle (3, 2);
          \fill[fill=green,fill opacity=0.3] (3, 0) rectangle (4, 2);
        	\fill[fill=magenta,fill opacity=0.5] (2, 4) rectangle (1, 3);
        	\draw (2, 4) -- (1, 4);
        	\draw (1, 4) -- (1, 3);
        	\draw (2, 3) -- (1, 3);
        	\fill[fill=magenta,fill opacity=0.5] (3, 4) rectangle (2, 3);
        	\draw (3, 4) -- (3 ,3);
        	\draw (3, 4) -- (2, 4);
        	\fill[fill=magenta,fill opacity=0.5] (3, 3) rectangle (2, 2);
        	\draw (2, 3) -- (2, 2);
        	\draw (3, 2) -- (2, 2);
        	\fill[fill=magenta,fill opacity=0.5] (4, 3) rectangle (3, 2);
        	\draw (4, 3) -- (4 ,2);
        	\draw (4, 3) -- (3, 3);
        	\draw (4, 2) -- (3, 2);
        \fill[pattern=north west lines, pattern color=green] (3, 0) rectangle (4, 2);
        \fill[pattern=north west lines, pattern color=green] (2, 1) rectangle (3, 2);
        \fill[pattern=north west lines, pattern color=green] (0, 3) rectangle (1, 4);
        \foreach \x in {0,...,5}{
		      \node[gray, above] at (6-0.5-\x,6) {\Large $x_{\x}$};
            \node[gray, right] at (6.5, 6-\x,) {\Large $y_{\x}$};
	    }
        \draw (7,0.8)  node {$\downarrow$};
        \draw (7,1.2)  node {$y$};
        \draw (1.0,6.8)  node {$\leftarrow$ $x$};
        \end{tikzpicture}
        }
        \caption{}
        \label{fig:s_shaped_hor}
    \end{subfigure}
    \caption{Design space for Tiles of the highest efficiency}
	\label{fig:high_class_design_space}
\end{figure}
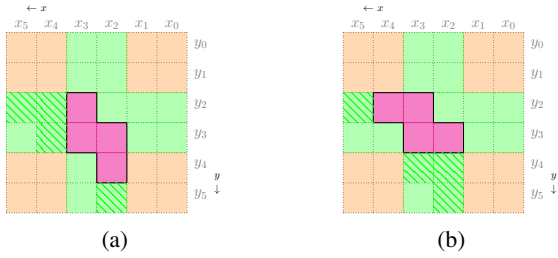

Note that the search has to be repeated for the four combinations of the signedness of the input vectors $X,Y$, which in each case can lead to yielding different efficiencies for a particular pattern.

\subsection{Identified Structures}
The search yielded structures with a multitude of efficiency levels up to $E=1.087$. Examples of tiles from the efficiency classes above $1.0$ are given in \figurename~\ref{fig:4x4search_results}. 
As the tiles with the highest efficiency by themselves are sufficient to generate complete tilings, we focus on those variants that generate overall efficient designs.
In total 31 structures were found within the highest efficiency class. At first glance these might appear to contain some randomness (see bottom row of \figurename~\ref{fig:4x4search_results}), their structure actually follows a few simple geometric design patterns. As shown in \figurename~\ref{fig:high_class_design_space}, the tiles can be broken down into in its most generalized form into a horizontal or vertical s-shaped element supplemented by an associated $1\times 1$ element which can be placed in the green area, such that an additional independent partial product bit is generated. All tiles within the highest efficiency class can be derived according to this rule, where some are redundant in the sense that the considered structure is placed twice within a tile.
In terms of its properties the tiles cover 5 positions on the board, generate 4 partial product bits and hence require 2 6-LUTs, each subdivided in two 5-LUTs to be realized.
For example, consider the s-shaped element in \figurename~\ref{fig:s_shaped_hor}, which after simplification is described by 
\begin{flalign*}
r_0&=\overline{y_3}y_2x_3+y_2\overline{x_2}x_3+y_3\overline{y_2}x_2+y_3x_2\overline{x_3}\\
r_1&=y_3\overline{x_2}\,\overline{x_4}x_3+y_3x_2x_4x_3+\overline{y_3}y_2x_4+y_2x_4\overline{x_3}+y_3\overline{y_2}x_3\\
r_2&=y_3y_2x_4x_3+y_3y_2x_2x_3 .
\end{flalign*}
Outputs $r_1,r_2$ already use 5 input variables, but $r_0$ requires only $4$ inputs. Hence, another arbitrary input can be combined with one of these 4 inputs to produce another partial product on a 1$\times$1 tile.
So for example when placing the 1$\times$1-element at (3,5), the partial product
\begin{flalign*}
r_3&=y_5x_3
\end{flalign*}
can be computed without additional cost.
Note that in the green hatched area the 1$\times$1-element has to be tabulated separately to permit to map it into two 6-LUTs as the corresponding equation would otherwise depend on 6 variables.

\input{figs/irr_tile_shapes}

\subsection{Selecting the Tile Set for the Optimization}

As the the complete list of tiles from the 4$\times$4 search space yielded an unpractical number of different tiles to be included in the optimization of the tiling, the tile set has to be narrowed down to a selection of the most efficient tile variants. 
Accordingly, all tiles with an efficiency above $1.0$ were selected and cleaned for redundancies which could be described as a combination of different tiles, yielding a tile set of $339$ elements. 
Optimization experiments for all multiplier dimensions between 1$\times$1 and 16$\times$16 were conducted, for which the solver generated optimal solutions up to about 13$\times$14 multipliers within a timeout of 10\,h. Beyond this, the solutions were only feasible due to the complexity of the model due to the amount of tiles.
Within those experiments $72$ of the $339$ tiles were actually used. 
The experiments were subsequently repeated with the set of $72$ used tiles, of which $54$ were actually used in the next round. 
Now all the results were optimal within the timeout and improvements could be observed in the cases that were previously not proven to be optimal due to the timeout. 
When examining those $54$ tiles geometrically it became apparent, that repeating redundant elements are present in tiles from some of the efficiency classes, which could be substituted by another tile with higher efficiency and a \textit{helper tile}, while maintaining the same realization costs, as illustrated in \figurename~\ref{fig:helper_tile_decomposition}. 
\def \tilescale{1}
\def \tilescaleouter{0.3}
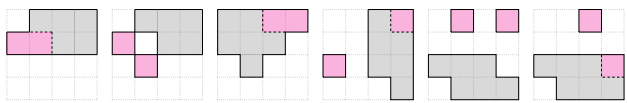
\begin{figure}[h]
\scalebox{\tilescaleouter}{
\begin{tikzpicture}[black,scale=\tilescale]
    \draw[step=1.0,dotted,thin,gray] (0,0) grid (4,4);
	\fill[fill=gray,fill opacity=0.3] (2, 4) rectangle (1, 3);
	\draw (2, 4) -- (1, 4);
	\draw (1, 4) -- (1, 3);
	\draw[dashed] (2, 3) -- (1, 3);
	\fill[fill=gray,fill opacity=0.3] (3, 4) rectangle (2, 3);
	\draw (3, 4) -- (2, 4);
	\fill[fill=gray,fill opacity=0.3] (4, 4) rectangle (3, 3);
	\draw (4, 4) -- (4 ,3);
	\draw (4, 4) -- (3, 4);
	\fill[fill=gray,fill opacity=0.3] (3, 3) rectangle (2, 2);
	\draw[dashed] (2, 3) -- (2, 2);
	\draw (3, 2) -- (2, 2);
	\fill[fill=gray,fill opacity=0.3] (4, 3) rectangle (3, 2);
	\draw (4, 3) -- (4 ,2);
	\draw (4, 2) -- (3, 2);
    \fill[fill=magenta,fill opacity=0.3] (0, 2) rectangle (2, 3);
    \draw (0, 3) -- (1, 3);
    \draw (0, 3) -- (0, 2);
    \draw (0, 2) -- (2, 2);
\end{tikzpicture}
}
\scalebox{\tilescaleouter}{
\begin{tikzpicture}[black,scale=\tilescale]
    \draw[step=1.0,dotted,thin,gray] (0,0) grid (4,4);
	\fill[fill=gray,fill opacity=0.3] (2, 4) rectangle (1, 3);
	\draw (2, 4) -- (1, 4);
	\draw (1, 4) -- (1, 3);
	\draw (2, 3) -- (1, 3);
	\fill[fill=gray,fill opacity=0.3] (3, 4) rectangle (2, 3);
	\draw (3, 4) -- (2, 4);
	\fill[fill=gray,fill opacity=0.3] (4, 4) rectangle (3, 3);
	\draw (4, 4) -- (4 ,3);
	\draw (4, 4) -- (3, 4);
	\fill[fill=gray,fill opacity=0.3] (3, 3) rectangle (2, 2);
	\draw (2, 3) -- (2, 2);
	\draw (3, 2) -- (2, 2);
	\fill[fill=gray,fill opacity=0.3] (4, 3) rectangle (3, 2);
	\draw (4, 3) -- (4 ,2);
	\draw (4, 2) -- (3, 2);
    \fill[fill=magenta,fill opacity=0.3] (0, 2) rectangle (1, 3);
    \fill[fill=magenta,fill opacity=0.3] (1, 1) rectangle (2, 2);
    \draw (0, 3) -- (1, 3) -- (1, 2) -- (0, 2) -- (0, 3);
    \draw (1, 2) -- (2, 2) -- (2, 1) -- (1, 1) -- (1, 2);
\end{tikzpicture}
}
\scalebox{\tilescaleouter}{
\begin{tikzpicture}[black,scale=\tilescale]
    \draw[step=1.0,dotted,thin,gray] (0,0) grid (4,4);
	\fill[fill=gray,fill opacity=0.3] (0, 3) rectangle (2, 4);
    \fill[fill=gray,fill opacity=0.3] (0, 2) rectangle (3, 3);
    \fill[fill=gray,fill opacity=0.3] (1, 1) rectangle (2, 2);
    \fill[fill=magenta,fill opacity=0.3] (2, 3) rectangle (4, 4);
	\draw (0, 4) -- (4, 4) -- (4, 3) -- (3, 3) -- (3, 2) -- (2, 2) -- (2, 1) -- (1, 1) -- (1, 2) -- (0, 2) -- (0, 4);
    \draw[dashed] (2, 4) -- (2, 3) -- (3, 3);
 \end{tikzpicture}
}
\scalebox{\tilescaleouter}{
\begin{tikzpicture}[black,scale=\tilescale]
    \draw[step=1.0,dotted,thin,gray] (0,0) grid (4,4);
	\fill[fill=gray,fill opacity=0.3] (2, 1) rectangle (3, 4);
    \fill[fill=gray,fill opacity=0.3] (3, 0) rectangle (4, 3);
    \fill[fill=magenta,fill opacity=0.3] (3, 3) rectangle (4, 4);
    \fill[fill=magenta,fill opacity=0.3] (0, 1) rectangle (1, 2);
	\draw (2, 4) -- (4, 4) -- (4, 0) -- (3, 0) -- (3, 1) -- (2, 1) -- (2, 4);
    \draw[dashed] (3, 4) -- (3, 3) -- (4, 3);
    \draw (0, 1) -- (1, 1) -- (1, 2) -- (0, 2) -- (0, 1);
 \end{tikzpicture}
}
\scalebox{\tilescaleouter}{
\begin{tikzpicture}[black,scale=\tilescale]
    \draw[step=1.0,dotted,thin,gray] (0,0) grid (4,4);
	\fill[fill=gray,fill opacity=0.3] (1, 0) rectangle (4, 1);
    \fill[fill=gray,fill opacity=0.3] (0, 1) rectangle (3, 2);
    \fill[fill=magenta,fill opacity=0.3] (3, 3) rectangle (4, 4);
    \fill[fill=magenta,fill opacity=0.3] (1, 3) rectangle (2, 4);
	\draw (0, 2) -- (3, 2) -- (3, 1) -- (4, 1) -- (4, 0) -- (1, 0) -- (1, 1) -- (0, 1) -- (0, 2);
    \draw (3, 3) -- (4, 3) -- (4, 4) -- (3, 4) -- (3, 3);
    \draw (1, 3) -- (2, 3) -- (2, 4) -- (1, 4) -- (1, 3);
 \end{tikzpicture}
}
\scalebox{\tilescaleouter}{
\begin{tikzpicture}[black,scale=\tilescale]
    \draw[step=1.0,dotted,thin,gray] (0,0) grid (4,4);
	\fill[fill=gray,fill opacity=0.3] (1, 0) rectangle (4, 1);
    \fill[fill=gray,fill opacity=0.3] (0, 1) rectangle (3, 2);
    \fill[fill=magenta,fill opacity=0.3] (3, 1) rectangle (4, 2);
    \fill[fill=magenta,fill opacity=0.3] (2, 3) rectangle (3, 4);
	\draw (0, 2) -- (4, 2) -- (4, 1) -- (4, 1) -- (4, 0) -- (1, 0) -- (1, 1) -- (0, 1) -- (0, 2);
    \draw (2, 3) -- (3, 3) -- (3, 4) -- (2, 4) -- (2, 3);
    \draw[dashed] (3, 2) -- (3, 1) -- (4, 1);
 \end{tikzpicture}
}
    \caption{Examples of the decomposition of tiles in combinations of other helper tiles}
	\label{fig:helper_tile_decomposition}
\end{figure}

It can be observed, that the introduction of the helper tiles does not only allow for the substitution of tiles from some efficiency classes in the remaining list, but also the creation of all tiles in the efficiency classes where the redundancies occur from the initial tile set ($E_t=1.013$ and $E_t=1.014$ in \figurename~\ref{fig:4x4search_results}). 
Most of the helper tiles cover two positions which are positioned relative to each other within the 4$\times$4 space, which resulted in a tile set of $55$ elements. 
The experiments showed slight improvements over the tile set without substitutions, and that $41$ of the $55$ tiles were actually used. 
Those $41$ tiles minus the 2$\times$3/3$\times$2 elements which had been introduced as a helper tile, but could be removed without changes to the optimization objective, form the basis for the final tile set that was  used for the synthesis experiments. 
The final tile set is shown in \figurename~\ref{fig:irr_tiles}, sorted by their number of uses (\#uses) in the experiment.

\section{Results}

\subsection{Experimental Setup}

The proposed incomplete tiles were added to the multiplier tiling framework of the open-source VHDL-code generator FloPoCo~\cite{dk24}.
For the experiments, the combined ILP-based optimization of the tiling and the compressor tree of \cite{Boettcher23} was used in conjunction with Gurobi 11 as the solver. The generated VHDL designs were subsequently synthesized and implemented with Vivado version 2022.1 for a Kintex 7 (xc7k70tfbv484-3) target FPGA.
\begin{table*}[ht!]
\centering
    \caption{LUT results from synthesis for the proposed incomplete multiplier tiling, as well as the rel. improvement in comparison to previous tiling \cite{Boettcher23} (in brackets) for $W_X\times W_Y$ multipliers}     
    \label{tb:impr_prev_tiling_t}
\begin{adjustbox}{max width=\textwidth}
\begin{tabular}
{@{\hspace{0.05cm}}l@{\hspace{0.05cm}}r@{\hspace{0.05cm}}c@{\hspace{0.2cm}}r@{\hspace{0.05cm}}c@{\hspace{0.2cm}}r@{\hspace{0.05cm}}c@{\hspace{0.2cm}}r@{\hspace{0.05cm}}c@{\hspace{0.2cm}}r@{\hspace{0.05cm}}c@{\hspace{0.2cm}}r@{\hspace{0.05cm}}c@{\hspace{0.2cm}}r@{\hspace{0.05cm}}c@{\hspace{0.2cm}}r@{\hspace{0.05cm}}c@{\hspace{0.2cm}}r@{\hspace{0.05cm}}c@{\hspace{0.2cm}}r@{\hspace{0.05cm}}c@{\hspace{0.2cm}}r@{\hspace{0.05cm}}c@{\hspace{0.2cm}}r@{\hspace{0.05cm}}c@{\hspace{0.2cm}}r@{\hspace{0.05cm}}c@{\hspace{0.2cm}}r@{\hspace{0.05cm}}c@{\hspace{0.2cm}}r@{\hspace{0.05cm}}c@{\hspace{0.2cm}}r@{\hspace{0.05cm}}c@{\hspace{-0.1cm}}}
\toprule
$W_Y$\textbackslash$W_X$&\multicolumn{2}{c}{1}&\multicolumn{2}{c}{2}&\multicolumn{2}{c}{3}&\multicolumn{2}{c}{4}&\multicolumn{2}{c}{5}&\multicolumn{2}{c}{6}&\multicolumn{2}{c}{7}&\multicolumn{2}{c}{8}&\multicolumn{2}{c}{9}&\multicolumn{2}{c}{10}&\multicolumn{2}{c}{11}&\multicolumn{2}{c}{12}&\multicolumn{2}{c}{13}&\multicolumn{2}{c}{14}&\multicolumn{2}{c}{15}&\multicolumn{2}{c}{16}\\ 
 \hline
1&\textbf{1}&(\textbf{0\%)}&1&(0\%)&2&(0\%)&2&(0\%)&3&(0\%)&3&(0\%)&4&(0\%)&4&(0\%)&5&(0\%)&5&(0\%)&6&(0\%)&6&(0\%)&7&(0\%)&7&(0\%)&8&(0\%)&8&(0\%)\\ 
2&1&(0\%)&\textbf{2}&\textbf{(0\%)}&3&(0\%)&5&(0\%)&6&(0\%)&7&(0\%)&8&(0\%)&9&(0\%)&10&(0\%)&11&(0\%)&12&(0\%)&13&(0\%)&14&(0\%)&15&(0\%)&16&(0\%)&17&(0\%)\\ 
3&2&(0\%)&3&(0\%)&\textbf{5}&\textbf{(0\%)}&9&(10.0\%)&12&(0\%)&14&(6.6\%)&16&(5.8\%)&18&(5.2\%)&21&(0\%)&23&(4.1\%)&25&(7.4\%)&28&(0\%)&30&(3.2\%)&32&(3.0\%)&34&(2.8\%)&35&(7.8\%)\\ 
4&2&(0\%)&5&(0\%)&9&(10.0\%)&\textbf{12}&\textbf{(7.6\%)}&15&(11.7\%)&18&(5.2\%)&20&(9.0\%)&23&(11.5\%)&26&(10.3\%)&28&(12.5\%)&31&(11.4\%)&34&(10.5\%)&37&(9.7\%)&39&(11.3\%)&42&(10.6\%)&45&(10.0\%)\\ 
5&3&(0\%)&6&(0\%)&12&(0\%)&14&(17.6\%)&\textbf{18}&\textbf{(10.0\%)}&23&(-4.5\%)&27&(3.5\%)&30&(3.2\%)&35&(0\%)&38&(2.5\%)&41&(2.3\%)&46&(3.8\%)&49&((0\%)&52&(1.8\%)&56&(0\%)&60&(0\%)\\ 
6&3&(0\%)&7&(0\%)&14&(6.6\%)&18&(-5.8\%)&22&(0\%)&\textbf{27}&\textbf{(0\%)}&34&(0\%)&38&(0\%)&42&(0\%)&46&(0\%)&50&(0\%)&54&(0\%)&58&(0\%)&62&(0\%)&66&(0\%)&70&(0\%)\\ 
7&4&(0\%)&8&(0\%)&16&(5.8\%)&20&(9.0\%)&27&(3.5\%)&34&(0\%)&\textbf{37}&\textbf{(9.7\%)}&42&(10.6\%)&47&(7.8\%)&52&(8.7\%)&58&(6.4\%)&63&(7.3\%)&68&(8.1\%)&73&(7.5\%)&78&(8.2\%)&83&(6.7\%)\\ 
8&4&(0\%)&9&(0\%)&17&(10.5\%)&23&(11.5\%)&30&(3.2\%)&38&(0\%)&42&(10.6\%)&\textbf{48}&\textbf{(7.6\%)}&54&(8.4\%)&61&(6.1\%)&67&(5.6\%)&73&(5.1\%)&79&(3.6\%)&85&(4.4\%)&91&(5.2\%)&97&(3.0\%)\\ 
9&5&(0\%)&10&(0\%)&21&(0\%)&26&(10.3\%)&36&(-2.8\%)&42&(0\%)&47&(6.0\%)&54&(10.0\%)&\textbf{61}&\textbf{(7.5\%)}&69&(5.4\%)&75&(6.2\%)&81&(5.8\%)&88&(4.3\%)&96&(4.0\%)&101&(4.7\%)&108&(5.2\%)\\ 
10&5&(0\%)&11&(0\%)&23&(4.1\%)&28&(12.5\%)&38&(2.5\%)&46&(0\%)&52&(10.3\%)&61&(6.1\%)&68&(6.8\%)&\textbf{75}&\textbf{(8.5\%)}&84&(5.6\%)&91&(5.2\%)&99&(3.8\%)&106&(3.6\%)&115&(1.7\%)&121&(2.4\%)\\ 
11&6&(0\%)&12&(0\%)&25&(3.8\%)&31&(11.4\%)&41&(2.3\%)&50&(0\%)&58&(6.4\%)&68&(4.2\%)&76&(5.0\%)&83&(6.7\%)&\textbf{93}&\textbf{(7.9\%)}&102&(6.4\%)&109&(7.6\%)&117&(7.1\%)&126&(5.2\%)&134&(5.6\%)\\ 
12&6&(0\%)&13&(0\%)&26&(7.1\%)&34&(10.5\%)&47&(-2.1\%)&54&(0\%)&65&(5.7\%)&73&(6.4\%)&81&(5.8\%)&91&(5.2\%)&102&(5.5\%)&\textbf{110}&\textbf{(6.7\%)}&120&(4.0\%)&128&(3.7\%)&137&(4.1\%)&146&(4.5\%)\\ 
13&7&(0\%)&14&(0\%)&30&(3.2\%)&37&(9.7\%)&49&(2.0\%)&58&(0\%)&68&(8.1\%)&79&(3.6\%)&88&(5.3\%)&98&(4.8\%)&109&(6.8\%)&120&(6.2\%)&\textbf{129}&\textbf{(6.5\%)}&140&(4.1\%)&149&(5.0\%)&160&(3.6\%)\\ 
14&7&(0\%)&15&(0\%)&32&(3.0\%)&39&(11.3\%)&52&(1.8\%)&62&(0\%)&73&(8.7\%)&85&(4.4\%)&94&(6.0\%)&106&(3.6\%)&118&(4.8\%)&128&(5.1\%)&139&(4.7\%)&\textbf{153}&\textbf{(3.1\%)}&162&(3.5\%)&173&(2.8\%)\\ 
15&8&(0\%)&16&(0\%)&34&(2.8\%)&42&(10.6\%)&56&(1.7\%)&66&(0\%)&78&(7.1\%)&92&(3.1\%)&101&(4.7\%)&114&(2.5\%)&126&(5.9\%)&137&(4.1\%)&149&(5.0\%)&163&(2.9\%)&\textbf{175}&\textbf{(5.4\%)}&186&(3.6\%)\\ 
16&8&(0\%)&17&(0\%)&35&(7.8\%)&45&(10.0\%)&60&(0\%)&70&(0\%)&84&(6.6\%)&97&(3.0\%)&108&(4.4\%)&121&(2.4\%)&135&(4.2\%)&146&(3.9\%)&159&(4.7\%)&174&(1.6\%)&186&(4.1\%)&\textbf{199}&\textbf{(2.9\%)}\\ 
\bottomrule
\end{tabular}
\end{adjustbox}
\end{table*}

\subsection{Comparison With Previous Tiling}

Table~\ref{tb:impr_prev_tiling_t} shows a comparison of synthesis experiment results for multipliers up to 16$\times$16 for the proposed incomplete multiplier tiles, as well as the relative improvement in comparison to the tiling using rectangular tiles \cite{Boettcher23}.
Using incomplete multiplier tiles offers improvements in terms of utilized LUTs in the vast majority of cases.
The reductions are up to 17.6\% with an average of 3.7\%.
The relative LUT reduction drops with increasing multiplier sizes but is still significant up to 16$\times$16.
Hence, incomplete multiplier tiles are most efficient for small multiplier sizes.
Some exceptions show no reductions, which are the very small (single LUT per output bit) multipliers and the ones which can be realized efficiently by the efficient 2$\times$\null$k$-tiles only.
For reference, the average improvement in the considered cases to the inferred multiplier (VHDL ``*'' operator) is 18.6\%. 

The objective usually closely matches the synthesis results, and as expected, are in every case better or equal than with the reference tile set. Although there is always some synthesis noise, like in the 4$\times$6-case where the reference result requires one LUT less, which results in a degradation of 5.8\%. 

\begin{figure}
\centering
\begin{subfigure}[b]{0.15\textwidth}
    \centering
    \includegraphics[width=1\textwidth]{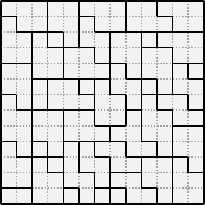}
    \caption{}
    \label{fig:13x13irr}
\end{subfigure}
\hspace{1cm}
\begin{subfigure}[b]{0.15\textwidth}
    \centering
    \includegraphics[width=1\textwidth]{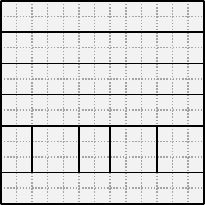}
    \caption{}
    \label{fig:13x13rect}
\end{subfigure}

\begin{subfigure}[b]{0.23\textwidth}
    \centering
    \input{./figs/svg/13x13/compression_13x13irr.tex}
    \caption{}
    \label{fig:13x13irr_comp}
\end{subfigure}
\begin{subfigure}[b]{0.23\textwidth}
    \centering
    \input{./figs/svg/13x13/compression_13x13rect.tex}
    \caption{}
    \label{fig:13x13rect_comp}
\end{subfigure}
\caption{Tiling and compression of a 13$\times$13-multiplier with (a),(c) incomplete and (b),(d) rectangular sub-multipliers}
\label{fig:13x13}
\end{figure}

Except large aspect ratio cases, the designs mostly consist of incomplete multiplier tiles although the 2$\times$\null$k$-multiplier-tile eventually becomes asymptotically more efficient towards $E_t=1.212$ with rising size and surpasses the efficiency of the incomplete tiles beyond 2$\times$12. The own realization costs per area of the proposed tiles is lower then of the 2$\times$\null$k$, but they entail higher compression costs. 
For example, the 13$\times$13 multiplier in \figurename~\ref{fig:13x13}a,c with incomplete tiles achieves overall lower realization cost of 130 LUT (objective) due to the less expensive tiling (70 LUT), even with higher compression costs (60 LUT), compared to the multiplier with rectangular tiles in \figurename~\ref{fig:13x13}b,d with 91 LUT (tiling) + 48 LUT (compression) = 139 LUT in total.   Nevertheless, the proposed tiles are also used in larger designs, where it is beneficial for the solver to substitute 2$\times$\null$k$-tiles for smaller tiles to overall generate less expensive designs for example by strategically rearranging the tiling to minimize the final adder length or effectively utilizing compressor inputs which would otherwise remain unused. 

\subsection{Comparison to Various Small State-of-the-Art Multipliers}

\begin{table}[t!]
    \caption{LUT comparison to previous rectangular tiling, fractal synthesis ported to AMD and Booth array}   
    \centering
    \begin{tabular}{lcccc}
        \toprule
        \textbf{Size}  &  \thead{\textbf{proposed}\\\textbf{tiling}\\ $\left[\text{LUT}\right]$} & \thead{\textbf{rectangular}\\\textbf{tiling\cite{Boettcher23}}\\ $\left[\text{LUT}\right]$} & \thead{\textbf{fractal}\\\textbf{synthesis\cite{lb18}} \\ $\left[\text{LUT}\right]$} & \thead{\textbf{Booth-}\\\textbf{Array \cite{Kumm15}}\\ $\left[\text{LUT}\right]$}\\
        \hline
        $3\times3$ & 5     &   5   & 6     &  9\\
        $4\times4$ & 12    &   13  & n.a.  & 17\\
        $5\times5$ & 18    &   20  & 20    & 23\\
        $6\times6$ & 27    &   27  & 32    & 35\\
        $7\times7$ & 37    &   41  & 42    & 39\\
        $8\times8$ & 48    &   52  & n.a.  & 51\\
        \bottomrule
    \end{tabular}  
    \label{tb:langhammer_vs_irr}
\end{table}

Table~\ref{tb:langhammer_vs_irr} shows a comparison between the proposed tiling using incomplete multipliers, the conventional tiling with rectangular tiles\cite{Boettcher23}, a Booth-Array-Multiplier \cite{Kumm15} and a multiplier optimized with the fractal synthesis approach \cite{Langhammer19,lb18}. 
Although the fractal synthesis approach was initially devised for the Intel architecture, it was ported to the AMD platform.
Note that it does not map efficiently to AMD devices, as an ALM (Intel 6-LUT) contains two full-adders, while AMDs 6-LUT contains only logic for one full-adder, resulting in almost twice as many LUTs on AMD. 
Hence, in our AMD port, we optimized the resources by using ternary addition for the compression.
The synthesis experiments were performed for unsigned multipliers.
It can be observed that for the examined small multipliers LUT savings can be achieved in many cases.
An average improvement to the previous tiling of 7.0\%, to fractal synthesis of 13.5\% and to the Booth array of 21.6\% could be achieved. 

\subsection{Comparison with Previous Truncated Tiling}

\begin{figure}
\centering
\begin{subfigure}[b]{0.15\textwidth}
    \centering
    \includegraphics[width=1\textwidth]{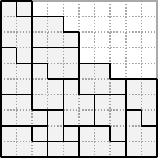}
    \caption{}
    \label{fig:10x10irr_trunc}
\end{subfigure}
\hspace{1cm}
\begin{subfigure}[b]{0.15\textwidth}
    \centering
    \includegraphics[width=1\textwidth]{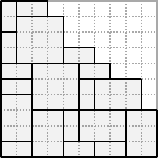}
    \caption{}
    \label{fig:10x10irr_rect}
\end{subfigure}
\caption{Tiling of a truncated 10$\times$10-multiplier with (a) incomplete and (b) rectangular sub-multipliers}
\label{fig:10x10trunc}
\end{figure}

For truncated multipliers, many LSB positions of the partial products do not need to be computed, resulting in tilings where the top-right corner of the board does not need to be covered (see \figurename~\ref{fig:10x10trunc}) \cite{Schulte93, Drane14, Boettcher21}. 
Here, the incomplete tiles also prove to be useful as their shape aligns well with staggered pattern between the realized MSBs and omitted LSBs. 

Table~\ref{tb:impr_prev_tiling_trunc} shows the LUT results from synthesis for different sizes up to 16$\times$16 where the output word size $W_\text{out}$ was set to the input word sizes $W_X=W_Y$.
Aside from the degradations of a single LUT due to synthesis noise in the $2\times2$ and $4\times4$ case, the proposed tiles offer more consistent improvements up to dimensions where the tilings become more similar to the solutions achieved with previous tile set, as the 2$\times$\null$k$-tiles become more efficient, and thus covers a larger share of the tiled area.
\figurename~\ref{fig:10x10trunc} shows a tiling of a faithfully-rounded truncated multiplier \cite{Boettcher21} with (a) incomplete and (b) rectangular sub-multipliers. 
While the tiling with rectangular tiles in \figurename~\ref{fig:10x10trunc}b requires 37 6-LUTs, only 31 6-LUT suffice for the tiling of \figurename~\ref{fig:10x10trunc}a, with compression costs of 28 and 29 6-LUT, respectively (objective). 

\begin{table}[t]
\centering
    \caption{LUT results for truncated multipliers for $W=W_y=W_y=W_\text{out}$}
    \label{tb:impr_prev_tiling_trunc}
\begin{adjustbox}{max width=\textwidth}
\begin{tabular}{lcccc}
\toprule
 $W$ & \textbf{proposed tiling} & \textbf{rectangular tiling} \cite{Boettcher23} & \textbf{improvement}\\
 \cmidrule(rl){1-4}
 1 &  1 &  1 & 0\%    \\ 
 2 &  1 &  1 & 0\%    \\ 
 3 &  3 &  7 & 57.1\% \\ 
 4 & 10 &  9 & -11.1\%\\ 
 5 & 15 & 17 & 11.7\% \\ 
 6 & 22 & 22 & 0\%    \\ 
 7 & 28 & 32 & 12.5\% \\ 
 8 & 37 & 40 & 7.5\%  \\
 9 & 48 & 51 & 5.8\%\\   
10 & 57 & 67 & 14.9\%\\  
11 & 68 & 78 & 12.8\%\\  
12 & 80 & 89 & 10.1\%\\  
13 & 96 & 104 & 7.6\%\\  
14 & 111 & 121 & 8.2\%\\ 
15 & 127 & 130 & 2.3\%\\ 
16 & 144 & 148 & 2.7\%\\   
\bottomrule
\end{tabular}
\end{adjustbox}
\vspace{-0.5\baselineskip}
\end{table}

\subsection{Packing Experiment}

To assess how well the proposed structures scale when used in more complex circuits, packing experiments were performed. Aside from LUT resources, the packing efficiency also considers the strain on routing resources which are harder to predict, as there are no detailed reports of the synthesis tools \cite{Langhammer19,lb18}. 
Following the methodology of previous work, the target device is iteratively filled with instances of the considered multiplier until place and route fails due to a lack of resources. 
The experiments were conducted for a combinatorial and pipelined (1 stage) unsigned $7\times7$ multiplier for the proposed method, the multiplier tiling with rectangular tiles \cite{Boettcher23}, the Vivado mult\_gen IP-core v12 \cite{Xilinx2015}, the inferred multiplier (VHDL ``*'' operator), the Booth-Array \cite{Kumm15,Walters14}, and a multiplier optimized with the fractal synthesis approach \cite{lb18}, ported to the AMD architecture. 
The experiments for the critical path delay (CPD) were performed with implementations of the tested multipliers sandwiched between registers, to ensure realistic conditions, similar as they would be when integrated in a larger design. Predictions for the theoretical number of instances which should fit into the FPGA were performed based on the synthesis results of a single instance, to evaluate if discrepancies arise compared to the actual number of placed instances the due to properties of the designs which are preventing fitment.

\begin{table}[t]
\caption{Packing experiment results for a $7\times7$ multiplier with various implementations}
\centering
\begin{adjustbox}{max width=0.48\textwidth}
\setlength\tabcolsep{3pt} 
\begin{tabular}{l l c C{0.8cm} C{0.8cm} C{0.8cm} C{0.8cm} C{0.8cm} C{0.8cm} C{0.8cm}}
\toprule
&\textbf{Type} & \multicolumn{2}{c}{\textbf{single mult.}} & \multicolumn{2}{c}{\textbf{\#mult/FPGA}} & \multicolumn{2}{C{1.9cm}}{\textbf{Utilization [\%]}}\\
\cmidrule(rl){3-4} \cmidrule(rl){5-6} \cmidrule(rl){7-8}
&             & \textbf{\#LUTs} & \textbf{CPD [ns]} & \textbf{theory} & \textbf{actual} & \textbf{Slice} & \textbf{LUT}\\
\midrule
\multirow{8}{*}{\rotatebox{90}{\bf combinatorial}}
&Proposed Tiling                         &     36 &     4.5  &   1138 &   1024 & 100.0 & 89.9 \\
&Rectangular Tiling \cite{Boettcher23}   &     42 &     4.8  &    976 &    964 & 100.0 & 97.6 \\
&Xilinx IP speed opt. v12                &     58 &     3.7  &    706 &    559 &  98.6 & 79.0 \\
&Xilinx IP area opt. v12                 &     79 &     4.2  &    518 &    471 & 100.0 & 90.8 \\
&Inferred multiplier                     &     51 &     3.8  &    803 &    713 &  99.9 & 87.9 \\
&Booth-Array \cite{Kumm15}               &     39 &     3.9  &   1051 &    841 &  98.5 & 80.0 \\
&Booth-Array \cite{Walters14}               &     32 &     3.8  &   1281 &    921 &  98.8 & 71.9 \\
&Fractal Synthesis \cite{lb18}           &     38 &     3.7  &   1078 &    822 &  99.6 & 76.1 \\
\midrule
\multirow{7}{*}{\rotatebox{90}{\bf pipelined}}
&Proposed Tiling                         &     36 &     3.0  &   1138 &   1024 & 100.0 & 89.9 \\
&Rectangular Tiling \cite{Boettcher23}   &     41 &     3.2  &   1000 &    759 & 100.0 & 75.9 \\
&Xilinx IP speed opt. v12                &     58 &     3.3  &    706 &    560 & 100.0 & 79.2 \\
&Xilinx IP area opt. v12                 &     77 &     3.2  &    532 &    422 &  99.9 & 79.3 \\
&Inferred multiplier                     &     51 &     3.6  &    803 &    713 & 100.0 & 87.8 \\
&Booth-Array \cite{Walters14}            &     32 &     2.6  &   1281 &    919 &  98.9 & 71.7 \\
&Fractal Synthesis \cite{lb18}           &     38 &     2.5  &   1078 &    819 &  99.4 & 75.9 \\
\bottomrule
\end{tabular}
\end{adjustbox}
\label{table:packing}
\vspace{-0.5\baselineskip}
\end{table}

The results in Table~\ref{table:packing} show, that the proposed method allows to fit the most instances to the device, although it has among the highest packing density within the slices. 
It can be observed that none of the evaluated methods possess properties that would prevent the utilization of vast majority of the slices. 
Unexpectedly, the Vivado IP core generator with the ``speed'' goal produces more compact multipliers than with the ``area'' goal, which confirms the finding in \cite{Kumm15}.

The packing result of \cite{lb18} is worse than expected from their experiment which might be due to our AMD-port and modifications.
It can generally be observed, that implementations that heavily rely on the fast-carry-chain achieve a lower LUT utilization, as residual LUTs within a slice can not be integrated in a different carry-chain.
As expected, the proposed tiling has a lower LUT utilization ($89.9$\%) compared to the rectangular tiling ($97.6$\%) due to the higher use of LUT inputs and outputs.
However, the low LUT usage per multiplier still dominates and leads to a significant increase of multiplier instances on the device.

\section{Conclusion}
The results show, that using incomplete sub-multipliers in the design of multipliers provides advantages in terms of resource utilization in comparison to previous methods.
They are especially beneficial for small multipliers as required for current AI inference tasks.
Although the experiments were performed for the AMD platform, the idea is applicable to all modern FPGAs. 
It is demonstrated, that the incomplete tiles can be used effectively with the multiplier tiling methodology to form complete multipliers using the combined ILP optimization approach for partial product generation and compressor tree design. As shown by the packing experiment, the designs can be used in complex circuits without particular limitations, so the advantages can directly translate to a higher overall arithmetic density in application like AI inference and DSP tasks.

\section{Acknowledgements}

The authors would like to thank Killian Klug for the help in implementing and porting previous work on multipliers.

\balance

\bibliographystyle{IEEEtran}
\bibliography{IEEEabrv,bibliography}

\end{document}